# Hydrodynamics of Clustered Clouds: Drafting, Survival, Condensation, and Ablation

M. Elliott Williams ⬤ and Robin L. Shelton ⬤

Department of Physics and Astronomy and Center for Simulational Physics, University of Georgia, 1003 Cedar St., Athens, GA 30602-2451, USA
elliott.williams14@gmail.com, rls@physast.uga.edu

## Abstract

For et al., who catalogued Magellanic Stream (MS) clouds, suggested that there is substantial large-scale turbulence in the MS. Here we follow up with a series of FLASH simulations that model the hydrodynamic effects that clouds have on each other. The suite of simulations includes a range of cloud separation distances and densities. The ambient conditions are similar to those surrounding the MS but also relevant to the circumgalactic medium and intergalactic medium. Ten simulations are presented, eight of which model clustered clouds and two of which model isolated clouds. The isolated clouds are used as controls for comparison with the multicloud simulations. We find that if the clouds are initially near each other, then hydrodynamical drafting helps the trailing cloud to catch the leading cloud and mix together. We present the measured acceleration due to drafting and find that lower-density clouds in lower-density environments experience more acceleration due to drafting than their denser cohorts. We find that the clustering of clouds also increases the condensation of ambient material and affects longevity. We analyze the velocity dispersion of the clouds using a single component method and a multicomponent decomposition method. We find that the presence of a second cloud increases the velocity dispersion behind the trailing cloud at some times. We find that the velocity dispersion due to gas motion in our simulations is significantly less than the actual dispersion observed by For et al., indicating that the thermal component must dominate in the MS.

*Unified Astronomy Thesaurus concepts:* Hydrodynamical simulations (767); Magellanic Stream (991); High-velocity clouds (735); Astronomy data analysis (1858); Hydrodynamics (1963)

## 1. Introduction

Our galaxy, the Milky Way (MW), is surrounded by orbiting satellite galaxies. Among these are the Large and Small Magellanic Clouds. As they travel through the MW's extended halo and circumgalactic medium, gas is tidally and ram pressure stripped from them, forming the Leading Arm (LA) in front of their path and the Magellanic Stream (MS) trailing behind (D'Onghia & Fox 2016).

The evolution of the LA and MS are of interest to astronomers because there is evidence that the diffuse gas that has been stripped off of them is able to fall onto the Galactic disk and cool enough to fuel star formation in the MW. For et al. (2014) published a catalog of 251 high-velocity clouds (HVCs) in the MS, many of which have head–tail morphologies, suggesting interaction with the MWs extended environment or other gas in the MS. For et al. (2014) noticed that the pointing direction of these HVCs is random, which they interpreted as an indication of strong turbulence. They suggested the shock cascade scenario (see Bland-Hawthorn et al. 2007) as a contributing process, where ablated cloud material colliding with downstream material generates turbulence (and Hα emission).

Typically, studies of clouds concentrate on individual clouds, but here we consider the effect of one cloud on another and how their interactions may contribute to the dynamics of clouds in the MS. One of the most important processes in cloud dynamics is the Bernoulli effect, in which a low-pressure pocket forms behind the leading object and pulls the trailing object toward it. While in that pocket, the trailing cloud experiences less ram pressure, allowing it to approach the leading cloud. A common term for this effect is drafting. Any trailing clouds are somewhat protected by the leading cloud from the drag force associated with traveling through the medium. In this scenario, the leading cloud also benefits because the low-pressure pocket behind it will be disrupted by the trailing cloud. Without the trailing cloud, this pocket pulls back on the leading cloud, restricting its movement through the medium.

We take a closer look at the processes of drafting and turbulence via simulations. We run numerical simulations of clouds in the MS using the University of Chicago's FLASH software (Fryxell et al. 2000). In order to examine the effects of drafting on cloud survival, we simulate cases that have two clouds, where one trails behind the other, and we simulate cases that have one cloud as a control. We have created velocity dispersion maps from the FLASH simulation data to visualize turbulence. We compare these generated maps with 21 cm observations. We use multiple methods to analyze the velocity dispersion in the simulations and discuss the implications of each method in relation to observational results. We also analyze the condensation and evaporation of material on the clouds throughout the simulations by tracking the mass of the clouds.

In Section 2, we describe the setup and initial parameters for the suite of 10 simulations, as well as the methods for calculating the acceleration, condensation, and velocity dispersion. In Section 3, we report the results of the simulations and analyses. These results include plots of the cloud positions as a function of time, plots of the net acceleration of the clouds due to drafting, analyses of the condensation and evaporation processes acting due to the presence of the second cloud, and analyses of the velocity dispersion maps. Lastly, we conclude in Section 4 with our interpretations of the aforementioned

**Table 1**
Simulation Models: Initial Parameters

| Simulation Name | $n_{amb}$ (cm$^{-3}$) | Initial Separation (pc) |
|---|---|---|
| dual_2rlow | $1.0 \times 10^{-4}$ | 600$\hat{z}$ |
| dual_4rlow | " | 1200$\hat{z}$ |
| dual_4r1olow | " | 1200$\hat{z} - 300\hat{x}$ |
| dual_8rlow | " | 2400$\hat{z}$ |
| sing_low | " | N/A |
| dual_2r | $1.0 \times 10^{-3}$ | 600$\hat{z}$ |
| dual_4r | " | 1200$\hat{z}$ |
| dual_4r1o | " | 1200$\hat{z} - 300\hat{x}$ |
| dual_8r | " | 2400$\hat{z}$ |
| sing | " | N/A |

**Note.** The ambient material has $T_{amb} = 2.0 \times 10^6$ K, and the cloud material has a core temperature of $T_{cl} = 2.0 \times 10^4$ K. The clouds are initially separated in the domain by the vectors in the Initial Separation column.
For all models, the initial cloud radius is 300 pc, and the wind tunnel speed is 140 km s$^{-1}$.

results, that clustering of clouds plays a role in cloud survival and distance traveled.

## 2. Methods

### 2.1. Simulations

We ran 10 simulations using the hydrodynamical simulation software FLASH v4.3 from the University of Chicago (Fryxell et al. 2000). Radiative cooling was modeled with the *m-05.cie* cooling curve from Sutherland & Dopita (1993). This curve approximates the gas as being in collisional ionizational equilibrium and having a metallicity of $10^{-0.5}$, which is between that of the background gas and the gas in the initial modeled clouds.

A summary of the initial parameters of these simulations is in Table 1. All simulations have a Cartesian coordinate system domain with dimensions of 2.4 kpc × 1.2 kpc × 12 kpc for the *x*-, *y*-, and *z*-directions, respectively. In the simulations, the *x–z* plane has reflection boundary conditions. Thus, a hemispherical portion of a cloud situated with its center intersecting the *x–z* plane will behave as a full cloud centered along the same plane. The clouds were modeled as half clouds. With this setup and assumption, we were able to save on computational resources. The cloud and ambient medium initially move relative to each other at a speed of 140 km s$^{-1}$ in the *z*-direction (a typical speed of HVCs in the tail of the MS measured in the Galactic standard of rest, GSR; Westmeier 2017). This was accomplished by moving the ambient medium past the cloud and through the domain using FLASH's wind tunnel simulation setup. Using the wind tunnel allowed the clouds to remain in the domain much longer than if the simulations were run in the halo reference frame.

Although approximations are made, the initial parameters of the simulations were chosen to mimic a typical section of the tail of the MS. The ambient material was an isotropic low-density plasma intended to mimic the environment of the tail of the MS, namely, the low-density extended MW halo and circumgalactic medium. Observations of O VI and O VII lines (Sembach et al. 2003; Bregman & Lloyd-Davies 2007), ram pressure stripping (Grcevich & Putman 2009), and simulations (Gatto et al. 2013) indicate a halo density of $0.1-5 \times 10^{-4}$ cm$^{-3}$ at the location of the MS, with more typical values on the lower end. Evidently, there is a large amount of uncertainty about this value. To help constrain the parameter space and explore the possible scenarios, we included models with ambient densities of $n_{amb} = 1.0 \times 10^{-3}$ and $1.0 \times 10^{-4}$ cm$^{-3}$. Using an adjusted mean of 0°.33 for the angular radii of clouds in the For et al. (2014) catalog and assuming a distance of 55 kpc to the MS section we want to simulate, an initial cloud radius of 300 pc was chosen. Cloud and ambient temperatures were chosen based on observations done by For et al. (2014) and Henley et al. (2014), and the wind tunnel speed was chosen based on H I observations from Putman et al. (2012).

The initial cloud density was set so that the cloud would have a temperature of $T_{cl} = 2 \times 10^4$ K and a density that was large enough for the cloud to be in hydrostatic balance with the ambient material at the beginning of the simulation. For the five simulations having $n_{amb} = 1.0 \times 10^{-4}$ cm$^{-3}$, this resulted in $n_{cl} = 1.0 \times 10^{-2}$ cm$^{-3}$. The other five simulations had 10 times larger ambient and cloud densities. To prevent a discontinuity shock between the cloud and ambient medium at the beginning of each simulation, we smoothed the density and temperature profile of each cloud so that it follows the equation

$$n(r) = -0.5(n_{cl} - n_{amb})$$
$$\times \tanh\left(\frac{r - 150 \text{ pc}}{20 \text{ pc}}\right) + 0.5(n_{cl} + n_{amb}). \quad (1)$$

Our use of smooth density profiles follows that of Vieser & Hensler (2007a), Heitsch & Putman (2009), and Kwak et al. (2011), who compared the effects of blunt and tapered profiles.

For each ambient density, we ran simulations with five different configurations. The first configuration contained only a single cloud and *sing* in its naming scheme. This configuration acted as the control. Three simulations were run with two identical clouds that start with separation distances of two, four, and eight times the radius of the cloud. The names of these simulations contain 2r, 4r, and 8r. The last simulation was run with the trailing cloud displaced by one radius length along the *x*-axis and trailing the leading cloud by four radii to provide a more realistic case of unaligned clouds (called 4r1o). The configurations for the suite of simulations that have $n_{amb} = 1.0 \times 10^{-4}$ and $n_{cl} = 1.0 \times 10^{-2}$ cm$^{-3}$ are displayed in Figure 1. The other suites of simulations have the same geometries as the pictured suite. In order to track the cloud material during postprocessing, we assigned it an oxygen metal abundance that was 0.001 of that of the initial ambient gas.

### 2.2. Cloud Tracking

In the drafting process, the leading cloud partially or fully obstructs the flow of gas onto the trailing cloud, consequently reducing the strength of the drag force felt by the trailing cloud. In addition, this obstruction creates a low-pressure region that can pull the trailing object toward it. The conditions for this drafting process occur in our wind tunnel simulations, which model clouds in the MS.

If drafting occurs in the simulations, we should expect to see the trailing cloud accelerate toward the leading cloud. However, the leading cloud is also affected. Without a cloud trailing closely behind it, the leading cloud would have more significant eddies at its tail, which would decelerate it relative to the ambient gas. By displacing these eddies, the trailing cloud mitigates their effects on the leading cloud. So, both clouds should experience a drafting-induced acceleration in the upwind direction that will have the effect of reducing the apparent drag force acting in the downwind direction. In the GSR, this translates to the clouds sustaining their velocities for a longer period of time than the single cloud, thus allowing clustered clouds in the MS to travel further. In the reference frame used in our simulations, the clouds were initially at rest relative to the domain but were accelerated in the direction of the ambient gas flow (toward $+\hat{z}$) by the drag force. If drafting occurs, then the clouds resist this acceleration, decreasing their net acceleration in the direction of the ambient gas flow.

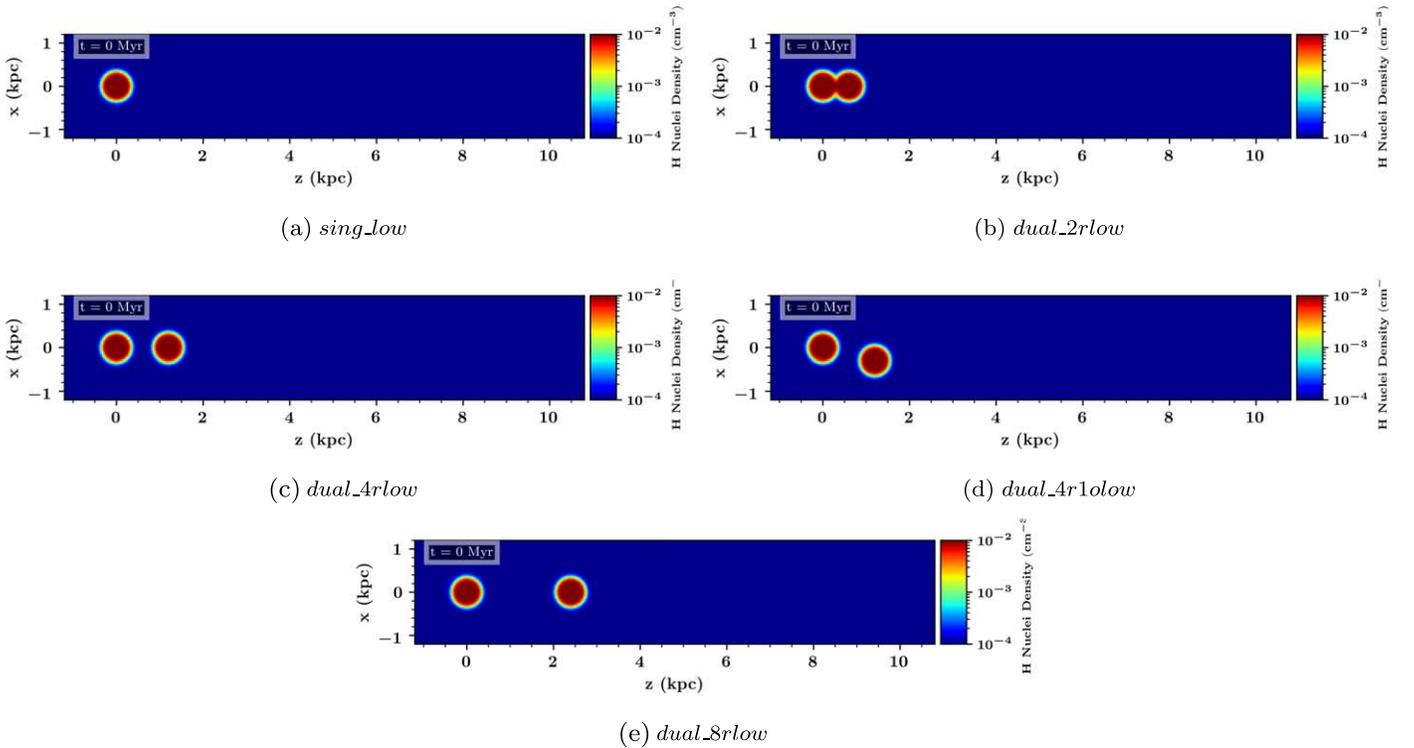

**Figure 1.** The initial configurations for the *sing_low*, *dual_2rlow*, *dual_4rlow*, *dual_4r1olow*, and *dual_8rlow* simulations are shown. The clouds in the other suites have the same locations. The color indicates the density in a slice through the domain, such that each maroon disk coincides with the location of a cloud in the initial setup.

We use metal abundance as an indicator for tracking the cloud cores through the domain. Over time, the cloud material and ambient material mix, complicating the determination of the effective cloud center. In order to deal with this problem, we calculate the fraction of each cell's mass that is derived from the original cloud. We use the cell's metal abundance to determine this fraction. Letting $\mathcal{M}_a \equiv$ the original metal abundance of the ambient gas, $\mathcal{M}_c \equiv$ the original metal abundance of the cloud, and $\mathcal{M}_i \equiv$ the metal abundance of any given cell, $i$, the fraction of mass in cell $i$ that came from the cloud is then

$$R_i = \frac{\mathcal{M}_a - \mathcal{M}_i}{\mathcal{M}_a - \mathcal{M}_c}. \tag{2}$$

We then calculate the distance of each cell in the $\hat{z}$-direction and multiply it by the fraction of cloud material in the cell. We then use mass weighting with metal abundance, meaning that we calculate the center of the cloud by weighting the cells with more cloud mass with higher magnitude. This helps us find the cells that contain the most material of the original cloud as it morphs and changes over the simulation. Letting $z_i$ be the distance in the $\hat{z}$-direction of cell $i$ and letting $m_i$ be the mass of that cell, the mass-weighted mean $\bar{z}$ of the cloud material is then

$$\bar{z} = \frac{\sum_i R_i m_i z_i}{\sum_i R_i m_i}. \tag{3}$$

Originally, the simulation grid contained cells of varying sizes because of FLASH's adaptive mesh refinement. However, performing Equation (3) requires that all cells have the same size. To have cells that are evenly sized, we created a fixed-resolution grid from the simulation grid by slicing large cells and combining small cells.

### 2.3. Turbulence Analysis

To investigate whether drafting is observably manifested by affecting the velocity dispersion, we created velocity dispersion maps of the simulational and observational data of the MS from the HI4PI survey (HI4PI Collaboration et al. 2016), a 21 cm all-sky survey that merges the Effelsberg-Bonn H I Survey (Kerp et al. 2011) and the Galactic All-Sky Survey (McClure-Griffiths et al. 2009). We have identified a region in the tail end of the MS that contains likely candidates for drafting.

To create the simulational velocity dispersion maps, we performed the analysis using both velocity and thermal data. Because of the random thermal motions of the particles, each cell does not contribute a discrete velocity value to the velocity dispersion but contributes a range of values to the overall distribution. The distribution of hydrogen velocities in each cell is modeled by a Gaussian curve of width equal to the thermal broadening component ($B_i$) and center equal to the bulk

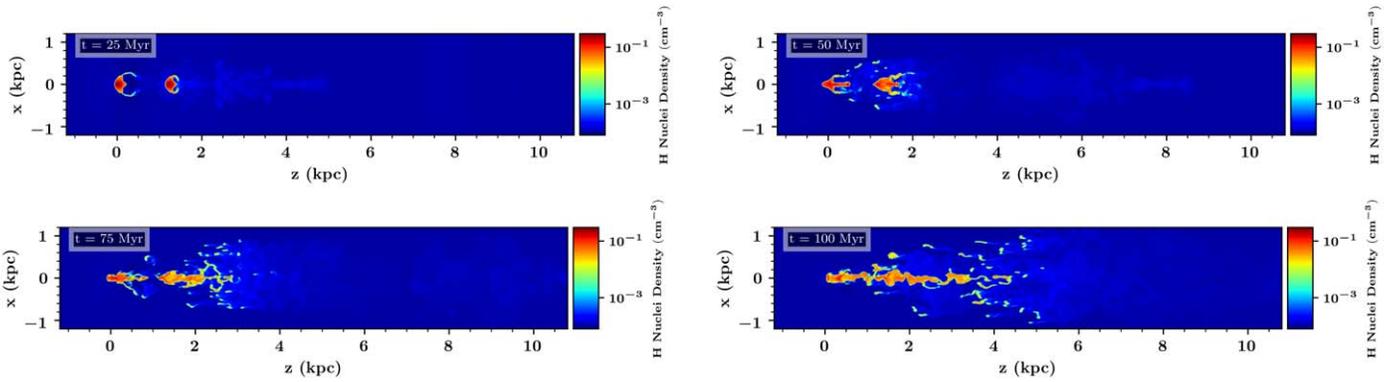

**Figure 2.** Plot of the density in the *x–z* plane at $y = 0$ in the *dual_4rlow* simulation at 25, 50, 75, and 100 Myr.

velocity of that cell. Here $B_i$ is calculated using the formula

$$B_i = \sqrt{\frac{2kT_i}{m_\mathrm{H}}}, \tag{4}$$

where $k$ is the Boltzmann constant, $T_i$ is the temperature of the cell $i$, and $m_\mathrm{H}$ is the mass of a hydrogen atom.

We sum the velocity contributions from each cell along a single line of sight to generate an overall velocity distribution that is used to calculate velocity dispersion. A predefined velocity range is split up into velocity bins of equal, predefined width. For each cell, a Gaussian distribution is created along the predefined velocity range, such that the center of the Gaussian is located at the cell's overall velocity in the direction of interest and the width of the Gaussian is equal to $B_i$. The Gaussian distribution is then allocated among the velocity bins so that the portion that lies in any given velocity bin $j$ is equal to the mass of the given cell that is traveling within that velocity bin range. The final weight of each bin is equal to the sum of all of the contributions that fall in that bin's range. The resulting distribution of mass as a function of velocity is normally a single bumpy curve resulting from the sum of the Gaussian contributions from each cell. The bumpiness arises from the nonuniformity of the centroid velocities. From it, we calculated the mass-weighted velocity dispersion along any given line of sight using

$$\sigma = \sqrt{\frac{\sum_j (v_j - \bar{v})^2 w_j}{\sum_i m_i}}, \tag{5}$$

where $w_j$ is the total mass per velocity bin for each velocity bin $j$, $v_j$ is the velocity of each velocity bin $j$, and $\bar{v}$ is the mass-weighted average velocity along the line of sight in the direction of viewing calculated with $\sum_j v_j w_j / \sum_i m_i$. We also adapted these formulae for FITS data cubes to compare our simulations to observational data. For comparison with the H I data, we filtered the simulational values so as to only include cells in which the temperature was between zero and $10^4$ K. We chose $10^4$ K as an upper limit because above this, it is unlikely to find H I, in accordance with the ionization balances from the cooling curve used in the simulation (Sutherland & Dopita [1993]).

## 3. Results

In all of the simulations, the clouds begin with spherical shapes but deform over time as Kelvin–Helmholtz instabilities stretch and pull material from them. Some of the ablated material collides with the second cloud or other ablated fragments, similar to the cascade scenario described by Bland-Hawthorn et al. ([2007]). In some cases, the leading cloud elongates to such an extent that it eventually collides with or envelops the second cloud. An example of cloud fragmentation, elongation, and cloud collision can be seen in the time series shown in Figure 2 of the *dual_4rlow* simulation at 25, 50, 75, and 100 Myr. The clouds in the *dual_4r*, *dual_8r*, and *dual_4r1o* configurations also stretch and develop before impact with the downstream cloud. In contrast, the clouds in the *dual_2r* simulation start with their edges touching, and the leading cloud ends up folding over the trailing cloud because they do not have enough time to develop on their own before the collision.

### 3.1. Tracking

In order to determine if drafting affects the net acceleration of the clouds, we calculated the average location of the clouds as a function of time using the procedures laid out in the methods section regarding the mass-weighted metal abundance tracking. We then calculated the distance a cloud has moved, $\mathcal{Z}$, from $\mathcal{Z} = \bar{z} - \bar{z}_{t=0}$, where $\bar{z}$ derives from Equation ([3]) and $\bar{z}_{t=0}$ is $\bar{z}$ at $t = 0$. As a result, each simulation's $\mathcal{Z}$ is equal to 0 pc at $t = 0$. This allows the curves to be compared more easily. Plots of the resulting $\mathcal{Z}$ as a function of time are presented in Figure 3.

We then fit $\mathcal{Z}$ as a function of time with a second-order polynomial of the form $\hat{\mathcal{Z}} = \frac{1}{2}at^2$. The variable $a$ corresponds to the acceleration of the cloud, which is due to the combined effects of drag, friction, and drafting. Because our initial cloud velocities with respect to the domain are $0\,\mathrm{km\,s^{-1}}$, we exclude a first-order term from the fitting. By the definition of $\mathcal{Z}$, the zeroth-order term at $t = 0$ is zero, so we exclude this term as well from the fits.

This method ignores any changes in cloud mass that occur during the simulations. Coefficient values can only reliably be compared between simulation sets with the same ambient conditions. To maintain pressure equilibrium, clouds in different ambient densities have difference masses; thus, their ablation behavior drastically deviates.

The fitting results in Table 2 show that the net acceleration (the positive acceleration due to the ram pressure and frictional forces combined with the negative acceleration due to the effects of drafting) of the two clouds in the low ambient density cases was less than the net acceleration of the single cloud. This

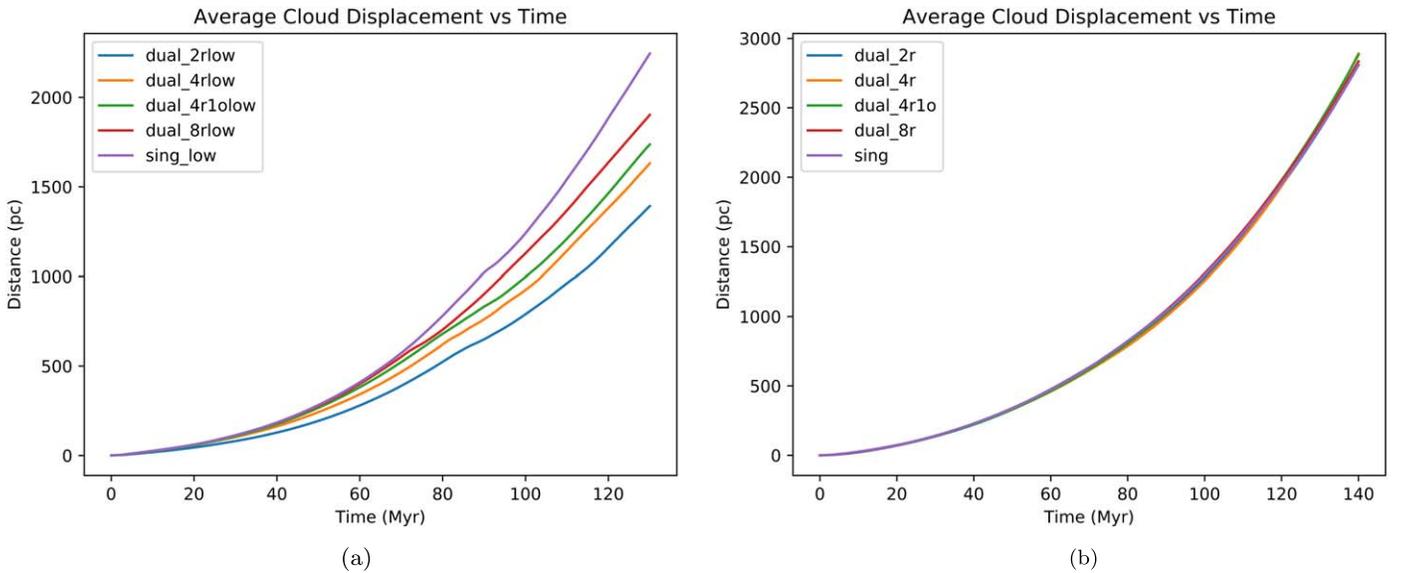

**Figure 3.** Plots of distance traveled by the clouds ($\mathcal{Z}$) due to drag, friction, and drafting. Left panel: low-density ($n_{\rm cl} = 1.0 \times 10^{-2}$ and $n_{\rm amb} = 1.0 \times 10^{-4}$ cm$^{-3}$) simulations. Right panel: high-density ($n_{\rm cl} = 1.0 \times 10^{-1}$ and $n_{\rm amb} = 1.0 \times 10^{-3}$ cm$^{-3}$) simulations.

**Table 2**
Acceleration

| Acceleration | |
| --- | --- |
| Simulation Name | $a$ (pc Myr$^{-2}$) |
| dual_2rlow | $0.1608 \pm 0.0002$ |
| dual_4rlow | $0.1904 \pm 0.0003$ |
| dual_4r1olow | $0.2039 \pm 0.0004$ |
| dual_8rlow | $0.2256 \pm 0.0002$ |
| sing_low | $0.2554 \pm 0.0008$ |
| dual_2r | $0.2697 \pm 0.0010$ |
| dual_4r | $0.2708 \pm 0.0013$ |
| dual_4r1o | $0.2752 \pm 0.0011$ |
| dual_8r | $0.2744 \pm 0.0009$ |
| sing | $0.2720 \pm 0.0008$ |

**Note.** Total acceleration experienced by the cloud material in each simulation, found by fitting the equation $\hat{\mathcal{Z}} = \frac{1}{2}at^2$ to the cloud material.

shows that drafting has a nonnegligible impact on the net force experienced by the clouds. To see that effect, consider that the downstream-directed acceleration due to the ambient medium's drag and frictional forces on the single cloud from the low-density simulation is $0.2554 \pm 0.0008$ pc Myr$^{-2}$. Assuming that the dual clouds experience the same drag and frictional force, the difference between the net acceleration in the *dual_2rlow* simulation ($0.1608 \pm 0.0002$ pc Myr$^{-2}$) and the net acceleration in the *sing_low* ($0.2554 \pm 0.0008$ pc Myr$^{-2}$) simulation approximately represents the acceleration caused by drafting in the dual-cloud simulation. The negative sign in the value for this acceleration (i.e., $-0.0946 \pm 0.0008$ pc Myr$^{-2}$) indicates that it is oriented in the upstream direction.

The low-density simulations showed greater signs of drafting than the high-density cases. As expected, the single cloud in the low-density simulations experienced the highest net acceleration. When we considered two clouds, there was a significant decrease in net acceleration as the initial positions of the clouds were brought closer together. The most extreme case, the

*dual_2rlow* simulation, experienced drafting equivalent to 37% of the net acceleration felt by the single cloud. For comparison, the least extreme case, the *dual_8rlow* simulation, experienced drafting equivalent to 12% of the net acceleration felt by the single cloud. The *dual_4rlow* simulation appeared to be slightly more advantageous than the *dual_4r1olow* setup, which agrees with previous studies of two-body drafting (Chatard & Wilson 2003; Westerweel et al. 2016).

The suite of high-density simulations did not show significant signs of drafting. To look for such signs, consider that the numerical values for acceleration listed in Table 2 are the result of multiple contributions, namely, those due to drag, friction, and drafting. Assuming that the contribution from drag is the same in all of the high-density simulations, the difference between the overall acceleration experienced by the single-cloud simulation and that for a corresponding dual-cloud simulation approximates the acceleration due to drafting plus any differences in frictional forces (e.g., shear) between the two simulations, which we assume are negligible. When this calculation was performed for the suite of low-density simulations, it yielded the appreciable drafting contributions noted in the previous paragraph. This calculation, however, yields little drafting contributions in the high-density suite of simulations. The greatest value for the drafting acceleration in the suite of high-density simulations is for the *dual_2r* simulation, and it is $0.0023 \pm 0.0013$ pc Myr$^{-2}$. In contrast with it, the *dual_2rlow* simulation experiences 41 times as much acceleration due to drafting, i.e., $0.0946 \pm 0.0008$ pc Myr$^{-2}$. The difference between the sizes of these accelerations is not due to the contribution from ram pressure.

We looked into a potential explanation for why the high-density suite did not show statistically significant signs of drafting and arrived at the following conclusions. We tested whether the acceleration due to ram pressure was greater in the high-density simulations than in the lower-density simulations. The acceleration due to ram pressure is a ratio of the force of ram pressure normalized by the cloud mass, and the ram pressure force is proportional to the ambient density. In both suites, the acceleration due to ram pressure is proportional to the ratio of (ambient density)/(cloud density), which is a factor of 0.01 in the high- and low-density suites of simulations. Therefore, the acceleration due to ram pressure is identical in both sets of simulations, refuting our potential explanation.

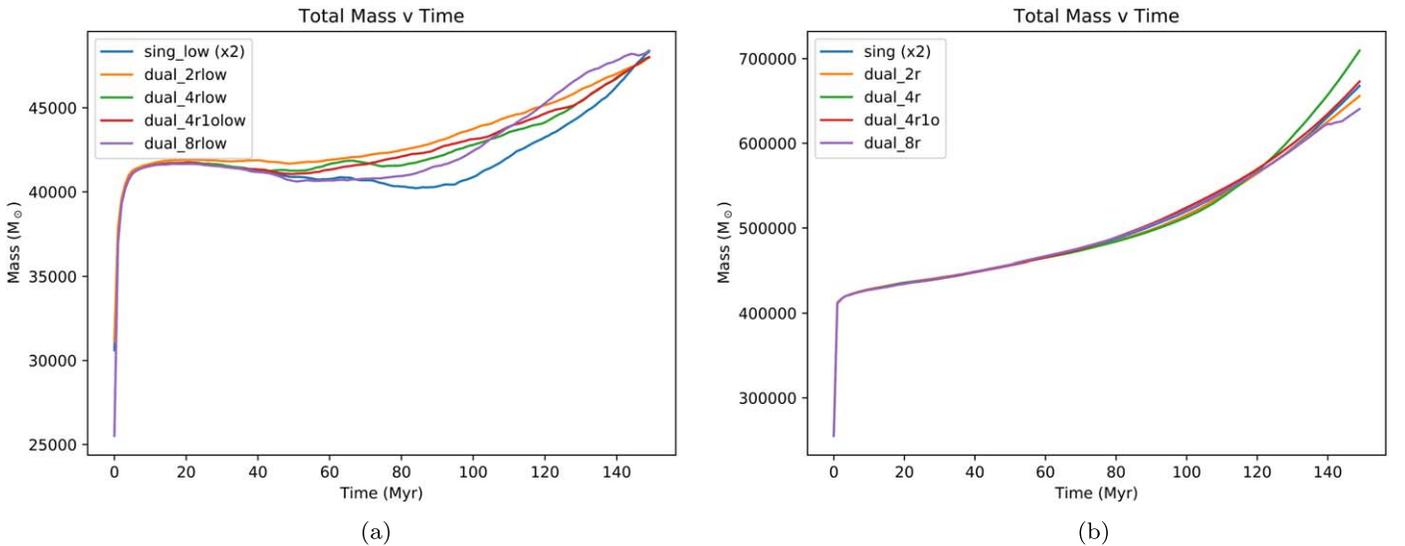

**Figure 4.** The mass of cool gas ($T \leqslant 2.5 \times 10^4$ K) in the domain is plotted as a function of time for the low- (left panel) and high-density simulations (right panel). The masses in the single-cloud simulations have been doubled for comparison with the dual-cloud simulations. These plots show the effects of condensation and evaporation on the total mass of cloud material over time.

### 3.2. Condensation

Over time, material from the ambient medium cools and condenses onto the clouds. The mass of cloud and condensed material is shown in Figure 4 as a function of time. Any ambient material that has cooled to a temperature of 25,000 K (similar to the initial cloud temperature) or below is considered to have condensed onto the cloud. We interpret any cloud material that has been heated or exists in the temperature regime above 25,000 K as evaporated material. Here we use a higher temperature threshold than in Section 2.3 to ensure that the entire cloud is captured in the tracking. Owing to the wind tunnel configuration of the simulations, ambient material is constantly flowing through the domain. At around 150 Myr, some of the ablated cloud material begins to flow out of the domain through the positive $z$ boundary. For this reason, we cut off the plots in Figure 4 at that time. Still, the clouds continue to condense mass beyond that point in time. For all simulations, the mass of the cloud(s) plus condensed material in our domains peaks at around 200–225 Myr at values of 727,000–1,042,000 $M_\odot$, equivalent to 1.88–2.7 times the initial dual-cloud mass of 386,000 $M_\odot$.

The low-density clouds experience short periods of condensation at the beginning of the simulations, but that is quickly overcome by ablation and evaporation beginning at ∼10 Myr. The extent of the ablation and evaporation is greatest in the single-cloud and *dual_8rlow* simulations but lesser in the *dual_4r1olow*, *dual_4rlow*, and *dual_2rlow* simulations. This is shown in Figure 4. The former simulations resume condensing more than evaporating between 50 and 90 Myr, but the latter simulations do so earlier. During this final condensation phase, the clouds accrete up to 43% of their initial masses before accretion is outdone by material leaving the domain. The peak in accreted mass occurs beyond the time range shown in Figure 4.

The best cases for drafting also proved to be the best cases for the retention of cloud material in the first approximately 90 Myr for the low-density cases. Situated just behind the leading cloud, the trailing cloud is protected from both turbulent motion that would contribute to Kelvin–Helmholtz instabilities (thereby allowing both of the clouds to travel through the domain with the least ablation) and ram pressure that would slow the cloud. Perhaps the retention of material by the clouds is in part caused by the protection from turbulent motion.

In contrast with the lower-density simulations, accretion always dominated over evaporation in the higher-density simulations. Immediately after these simulations started, accretion began to dominate over evaporation. This can be seen by the monotonic increase in mass in Figure 4, which requires that the processes causing the clouds to gain mass dominate over the processes that cause the clouds to lose mass. We assign the actions of condensation and evaporation to these processes. The trend continued even after material began to leave the domain at 150 Myr. Ultimately, the most closely spaced clouds gained more mass than their further-spaced cohorts. The *dual_2r* simulation packed on 170% more than its initial mass, while the *dual_8r* gained 88% of its initial mass before accretion was made irrelevant by the effect of material flowing off of the domain.

### 3.3. Velocity Dispersion

Figure 5 provides example dispersion maps of the *dual_4rlow* and *dual_4r* simulations showing the velocity dispersion in the $\hat{y}$-direction. In this figure, the leading clouds have been stretched so much that they have collided into the head of the second clouds, producing areas of high velocity dispersion. While the high velocity dispersion is due in part to the dynamics of the cloud material, most of the calculated dispersion is due to thermal broadening of the cloud gas, whose simulated temperature is $T \sim 10^4$ K. The edges of the cloud have higher temperatures than the cores and so show greater dispersion in these plots.

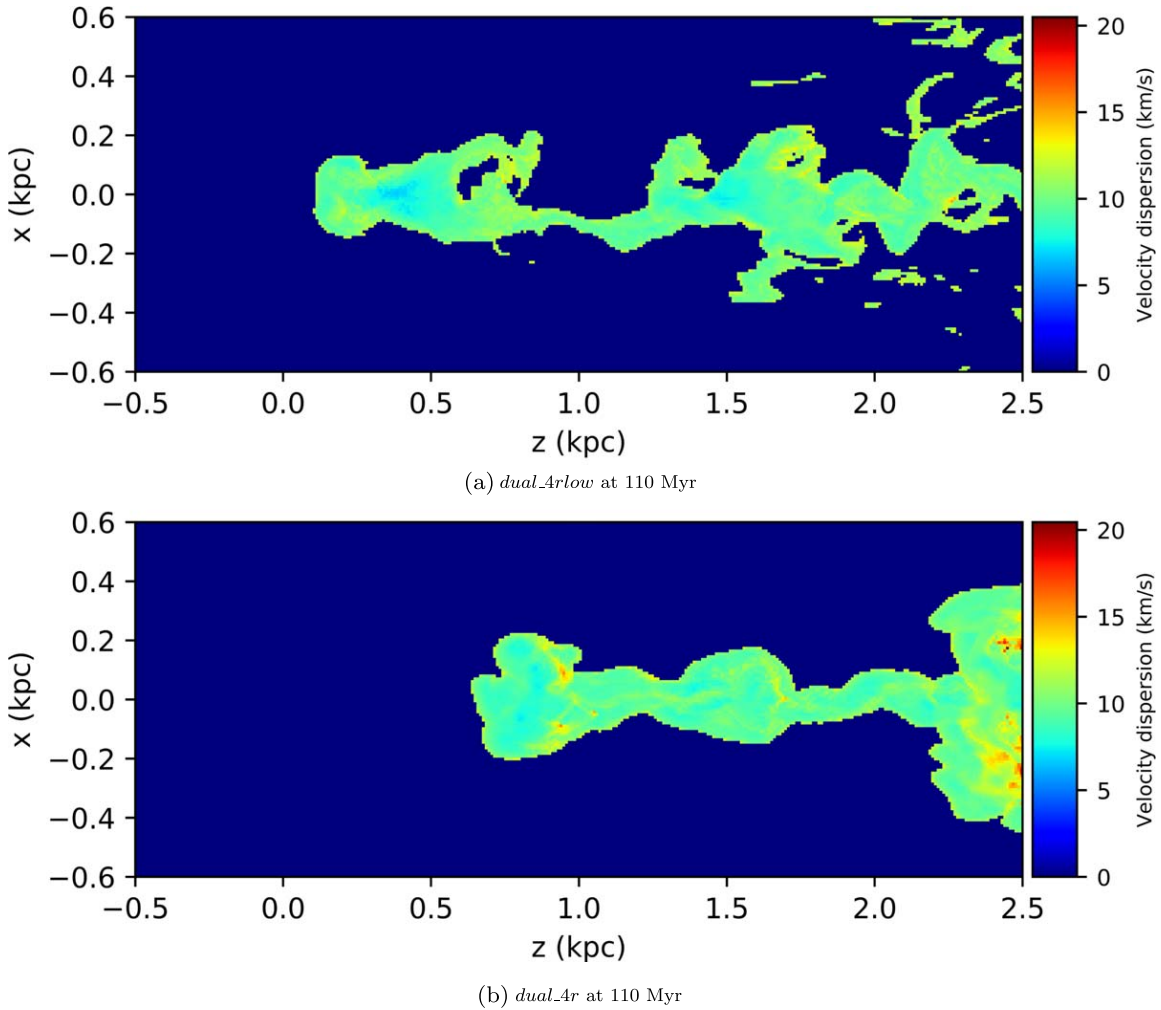

(a) *dual_4rlow* at 110 Myr

(b) *dual_4r* at 110 Myr

**Figure 5.** Comparison of velocity dispersion maps of the *dual_4rlow* and *dual_4r* simulations at 110 Myr.

Considering that many clouds in the MS and other complexes have lower temperatures, we next perform the same calculations under the assumption of lower cloud temperatures. For this, we calculated each cell's contribution to the thermal broadening as if the cell's temperature were 100 or 0 K. This is equivalent to substituting these temperatures for the temperatures of cloud material in the domain. This change decreases the thermal broadening contribution to the velocity dispersion. Dispersion maps were made in which the thermal broadening component used a temperature of either 100 or 0 K. This results in the maps in Figure 6 and later figures. Minimizing the thermal broadening component helps to reveal more detailed structures.

Figure 6 provides an example showing dispersion maps of selected low-density simulations with velocity dispersion in the $\hat{y}$-direction and the substituted cloud temperature of 0 K. This removes all effects of thermal broadening, leaving the dispersion due to bulk motion. Overall, the material in the single- and dual-cloud simulations has similar values for velocity dispersion. The median values in these plots of the single and dual clouds are $\sigma = 1.0$ and 1.31 km s$^{-1}$, respectively. The peaks are similar as well. The peak value for the dual-cloud simulation is $\sigma = 9.75$ km s$^{-1}$, while the peak for the single-cloud simulation is $\sigma = 9.64$ km s$^{-1}$. These values are not representative of the entire simulation throughout time; there is greater velocity dispersion at earlier times, owing to strong small-scale turbulence behind the low-pressure pocket of the trailing cloud. Nor are these values representative of the high-density simulations, which have much higher velocity dispersion values for similar time epochs and configurations. Velocity dispersion in the high-density cases is discussed in more detail later.

At 50 Myr, the peak value for the dual-cloud simulation is close to twice that of the single-cloud simulation ($\sigma = 15.72$ and 9.58 km s$^{-1}$, respectively). It is important to note that the majority of the high dispersion in the dual-cloud simulation at 50 Myr was behind the second cloud. This is due to the positioning of the second cloud in the low-pressure pocket behind the leading cloud. This region has strong eddies that generate turbulence where the flow trajectories around the clouds converge. While the signatures we anticipated (stronger velocity dispersion between the clouds) were not observed, we do observe a stronger velocity dispersion behind the second cloud. These characteristics are in line with other expectations about drafting.

When the temperature of the cloud material is set to 100 K, the resulting thermal broadening contribution to velocity dispersion creates a constant background level, or silhouette, of the clouds in the velocity dispersion maps. Figures 7 and 8 show these footprints. This trend is found for all simulations when we reset the temperature, regardless of ambient density, number of clouds, or time epoch observed.

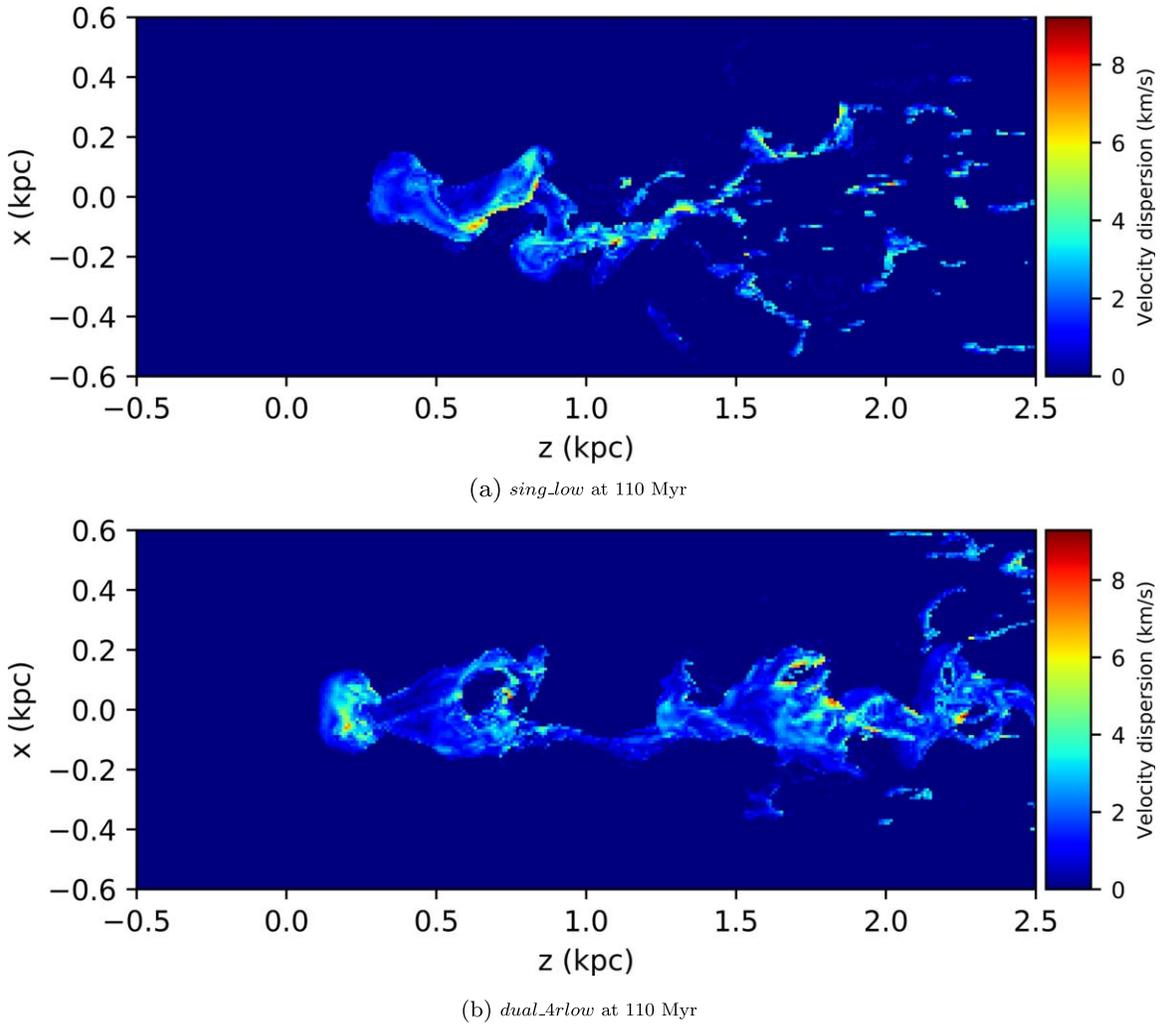

(a) *sing_low* at 110 Myr

(b) *dual_4rlow* at 110 Myr

**Figure 6.** Comparison of velocity dispersion maps of the *sing_low* and *dual_4rlow* simulations at 110 Myr, calculated without a thermal broadening component.

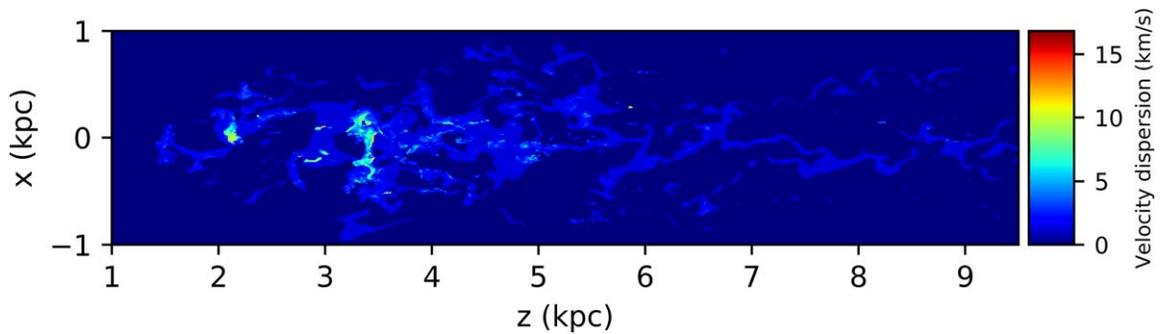

**Figure 7.** Velocity dispersion map of the *dual_4rlow* simulation at 200 Myr with the thermal broadening component set to a constant, resulting from a substituted temperature of 100 K.

The clouds tend to fragment, and the extent of the fragmentation builds up over time. Figure 9 shows an example: the *dual_4rlow* simulation at 200 Myr. Almost all of the small cloud fragments, with sizes in the range of ∼75–150 pc, tend to have high dispersion values. In the low-density simulations, the clouds fragment more than in the high-density simulations, which can be seen in Figure 5.

The range of dispersion values for the clouds matched closely to ones found in For et al. (2014). The majority of the clouds in the catalog have velocity dispersion values of 15–20 km s$^{-1}$ (the distribution follows a rough Gaussian shape with a mean equal to 15.7 km s$^{-1}$ and a standard deviation of 7 km s$^{-1}$). As noted in Section 2.3, we calculated the velocity dispersion for cloud material, which, for the sake of comparison with H I data, was assumed to be all material whose temperature was less than or equal to 10$^4$ K. At this temperature, thermal broadening accounts for $\sigma = 12.9$ km s$^{-1}$. Thus, the majority of the observed velocity dispersion can be attributed to thermal dispersion. Still, this adds another level of congruency between the simulations and the observed clouds in the catalog that is worth mentioning.

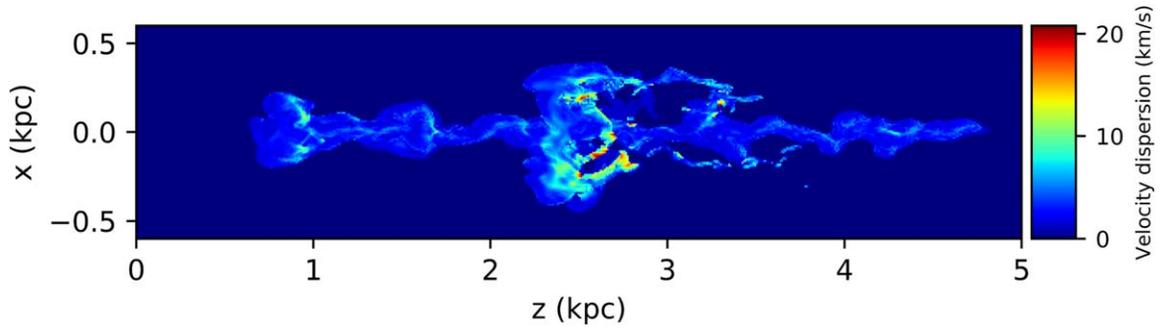

**Figure 8.** Velocity dispersion map of the *dual_4r* simulation at 110 Myr with the thermal broadening component set to a constant, resulting from a substituted temperature of 100 K.

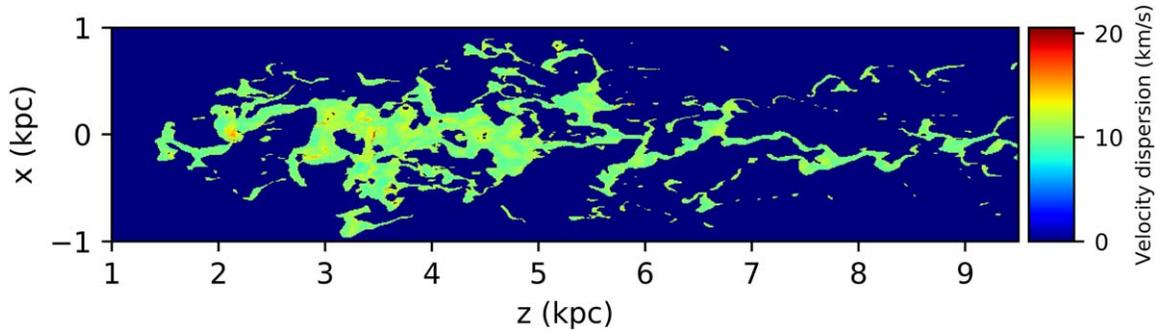

**Figure 9.** Velocity dispersion map of the *dual_4rlow* simulation at 200 Myr.

We observe relatively high velocity dispersion along the circumference of clouds in the cases where we calculate the thermal broadening directly from the simulational temperature (Figures 5 and 9). This trend of high velocity dispersion around the periphery is found in all simulations and is due to the hot gas from the ambient medium mixing with the cloud material at the boundary between the two. This is confirmed by looking at integrated temperature plots of the simulations. Considering that the clouds fragmented more in the low ambient density simulations, this effect is more visible in those plots because the fragments mix faster with the hot ambient medium. For confirmation, we tested the effect of reducing the cloud temperature to a uniform zero or 100 K and recalculating the velocity dispersion. With uniform cloud temperatures, the clouds cease to have a membrane of high velocity dispersion. The gas at the periphery is still relatively cool gas showing up in the dispersion maps, though, because our velocity dispersion calculations examined only material with a temperature less than or equal to 10,000 K in order to focus on material that would appear in H I 21 cm observations. Note that because this method of filtering involves temperature rather than species, the periphery effect is likely to show up to a lesser degree in H I 21 cm observations.

We were also able to produce velocity dispersion plots of the clouds along the line of sight of the wind tunnel. Figure 10 shows an example of this from the *dual_4r* simulation. Note that the velocity dispersion values are much higher than in the maps from the side view of the cloud (such as Figure 5). This is because the view includes multiple cloud fragments with their own velocities. Considering that overlapping clouds result in a higher-than-typical velocity dispersion, this suggests that regions of comparatively high velocity dispersion may be associated with overlap of clouds. Consider several clouds near each other that are observed and their velocity dispersions determined. The cores of clouds tend to be cooler than the extremities. Thus, for nonoverlapping clouds, the interiors of the clouds would not have a higher velocity dispersion than the outskirts. In contrast, when clouds overlap, the difference in bulk velocity contributes to the velocity dispersion, causing the overlapping regions to have a higher total velocity dispersion.

When our perspective shifts from the side view to the view along the direction of the flow, the effects of overlapping clouds on the velocity dispersion become apparent. For example, the side view of the *dual_4r* simulation seen in Figure 5(b) has a maximum velocity dispersion of $\sigma = 23.51$ km s$^{-1}$. The same simulation viewed from the perspective of the flow of the clouds in Figure 10 has a maximum value of $\sigma = 38.28$ km s$^{-1}$, or 1.63 times the amount seen in the side view. In the cases where the temperature for thermal broadening has been replaced, we see an even more pronounced effect. In the *dual_4r* simulation with a substituted temperature of 100 K shown in Figure 8, the maximum velocity dispersion is $\sigma = 20.79$ km s$^{-1}$. In contrast, the same simulation with the same substituted temperature, seen along the direction of cloud motion as shown in Figure 11, has a maximum velocity dispersion of $\sigma = 36.72$ km s$^{-1}$, or 1.77 times the amount seen in the side view (Figure 8).

Differences in cloud velocity centroids along the line of sight create larger velocity gradients than the motion within the clouds and therefore increase the velocity dispersion. This logic may also be applied to observational analyses when examining the velocity dispersion maps in conjunction with velocity centroid maps of H I observations. For example, when multiple clouds overlap along the light of sight and have different velocities, they create a spike in dispersion, as seen in Figure 12 with the *dual_4r1o* simulation. In the case where the clouds partially overlap along the line of sight in Figure 12, the maximum is roughly $\sigma = 25$ km s$^{-1}$. However, the maximum velocity dispersion is higher when the clouds overlap completely. An example of this is in Figure 10, where it peaks at $\sigma = 40$ km s$^{-1}$.

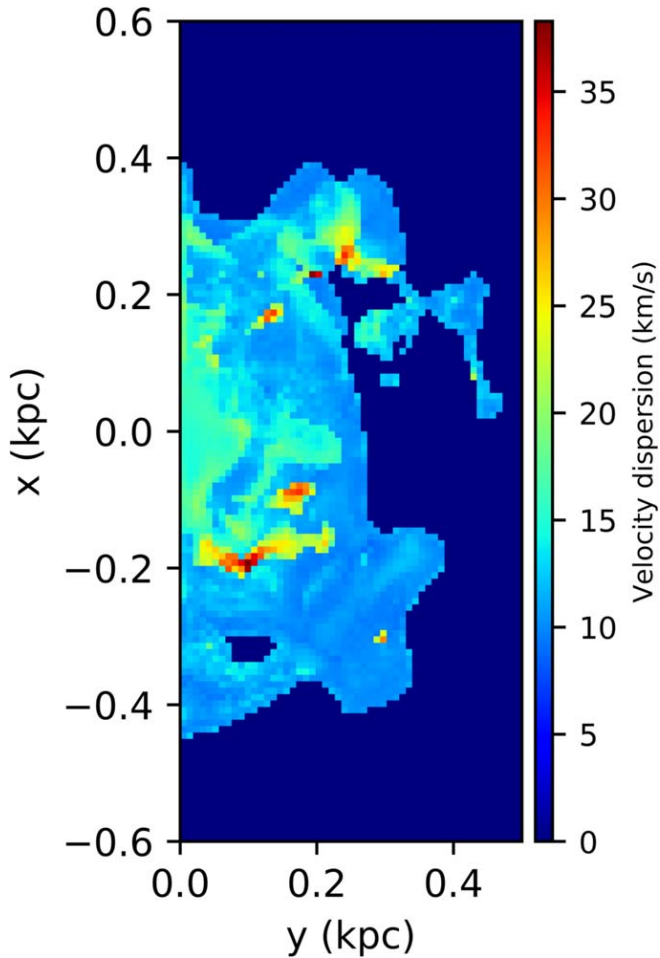

**Figure 10.** Velocity dispersion map of the *dual_4r* simulation at 110 Myr from the perspective along the direction of flow of the clouds.

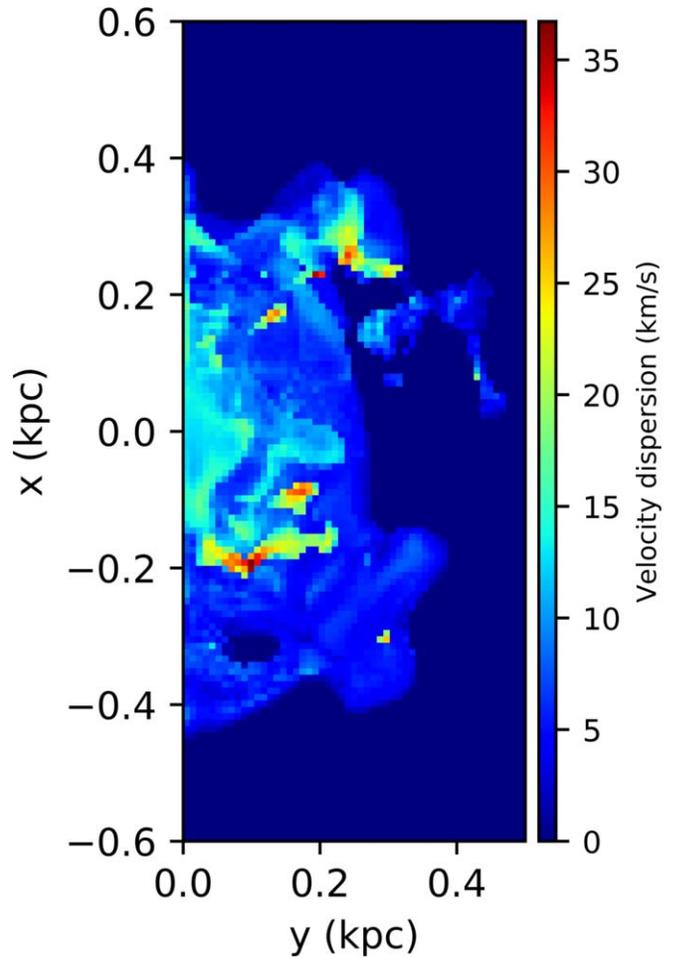

**Figure 11.** Velocity dispersion map of the *dual_4r* simulation at 110 Myr from the perspective along the direction of flow of the clouds. The thermal broadening component is set to a constant value equal to that in which the assumed temperature is 100 K.

In practice, the process of finding regions of overlapping clouds has two parts. Consider, for example, the H I4PI data (HI4PI Collaboration et al. 2016) for the MS within $l = (70.0, 80.0)$ and $b = (-50.0, -57.5)$, whose velocity dispersion and centroid maps are shown in Figure 13, overlaid with the H I column density contours. This segment of MS V (the fifth region of clouds along the MS identified by Mathewson et al. 1977) contains multiple cloudlets arranged advantageously for drafting because they are aligned parallel to the assumed motion of the MS. We have first identified a region centered at $(l = 74.4, b = -56.4)$ with an elongated shape in a slanted orientation that has a higher dispersion than the surrounding material. We corroborate this evidence with information from the velocity centroids map seen in Figure 13. In this region, we see a sharp gradient in centroid velocity. The centroid velocities are different on either side of the higher-dispersion region, suggesting that these are, in fact, two separate clouds.

In addition to using the velocity centroid and dispersion maps to identify multiple clouds along a line of sight and analyze the complex velocity structure of the simulations, we present another method: automated multicomponent Gaussian decomposition for simulations. The prior method for calculating velocity dispersion (using Equation (5)) calculates the dispersion due to all velocities observed along the line of sight. However, it is possible to model the line-of-sight velocity structure as a collection of overlapping Gaussian-shaped components in the event that there are multiple velocity components along the line of sight. It is possible to separate individual overlapping Gaussian components from the simulated emission line and fit them in order to approximate their individual dispersion components. Riener et al. (2019) developed a package called GAUSSPY+ for observational data that achieved this with remarkable results. We adopted the fitting method from Riener et al. (2019) to develop a similar codebase for simulational data that uses the velocity structure of the simulational data to identify and separate individual dispersion components.

A strength of this new methodology is that it makes it possible to determine the distribution of the strongest velocity components in velocity space. The resulting maps include a map of the strongest velocity dispersion components (e.g., Figure 14), a map of the centroids of the strongest velocity

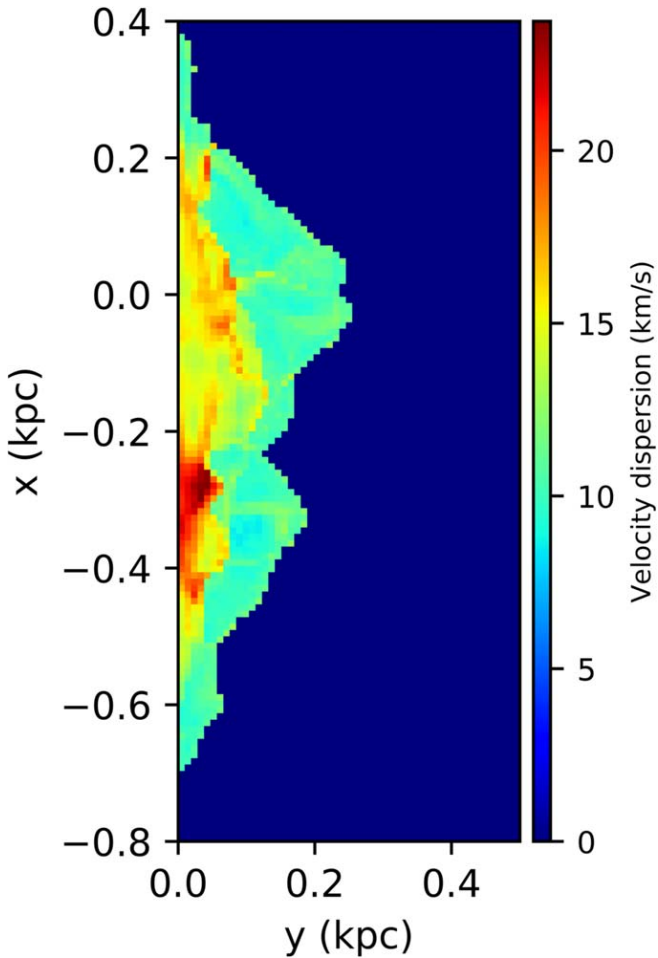

**Figure 12.** Velocity dispersion map of the *dual_4r1o* simulation at 110 Myr from the perspective along the direction of flow of the clouds.

The velocity dispersion in these simulations is the result of the thermal broadening and velocity gradients along the line of sight. Using the velocity gradients in the simulated domain to approximate the velocity gradients in observational data may make it possible to estimate the temperature of the observed material.

As an example of the above logic, we consider a small cloud in the fifth section of the MS. It is at $(l,\ b) = (76.0, -54.5)$ and shown in Figure 13. This cloud is not part of an overlapping conglomeration of clouds and is $\sim$950–1300 pc behind the cluster of clouds upstream from it, assuming that this part of the MS is $\sim$54–74 kpc from Earth (using the galactocentric distances to the MS from Brüns et al. 2005 and Jin & Lynden-Bell 2008 and the distance to Sgr A* from Abuter et al. 2019). Such a gap is approximately four times the radius of one of our simulated clouds, enabling us to compare its velocity dispersion with that in our *dual_4r* simulation. According to HI4PI Collaboration et al. (2016), the cloud has an H I column density of $6.3 \times 10^{18}\ \mathrm{cm}^{-2}$, which is 2.7 times the $5\sigma$ detection limit quoted in their observation methods. The median velocity dispersion in this direction is $\sigma = 9.66\ \mathrm{km\,s}^{-1}$. If the dispersion were entirely due to thermal broadening, this value would correspond to a temperature of $T = 5650$ K. However, a portion of the observed dispersion must be due to velocity gradients. If the observed cloudlet is similar in orientation and dynamics to our modeled *dual_4r* clouds at 110 Myr shown in Figure 17, then its median $\sigma$ due to velocity gradients alone is $\sigma = 2.39\ \mathrm{km\,s}^{-1}$. In that case, subtracting the velocity gradient contribution ($\sigma = 2.39\ \mathrm{km\,s}^{-1}$) in quadrature from the total velocity dispersion in the HI4PI clouds results in a thermal component of $\sigma_{\mathrm{thermal}} = 9.36\ \mathrm{km\,s}^{-1}$, corresponding to a temperature of $T = 5300$ K. Since thermal broadening accounts for nearly all of the velocity dispersion, the value of $\sigma^2 m_{\mathrm{H}}/2k$ is a good proxy for the gas temperature.

## 4. Discussion

We examined how clustering of clouds affects drafting, condensation/evaporation, and velocity dispersion as the clouds travel through low-density environments like that surrounding the MS. We performed a suite of simulations, concentrating on two sets of densities and several values of cloud separation distances.

As the clouds move through the surrounding medium, they experience ram pressure that decelerates them. However, multiple simulated clouds decelerate less than isolated clouds, indicating that drafting contributes an acceleration toward the upwind direction. This effect was strongest in our simulations of lower-density clouds ($n_{\mathrm{cl}} = 1.0 \times 10^{-2}\ \mathrm{cm}^{-3}$) in a lower-density medium ($n_{\mathrm{amb}} = 1.0 \times 10^{-4}\ \mathrm{cm}^{-3}$). In these models, the strength of drafting showed a negative correlation with the initial separation distance between the clouds, such that closer clouds showed greater signs of drafting. In the most extreme case, drafting was about one-third as strong as ram pressure. Signs of drafting were far less consistent in the $n_{\mathrm{cl}} = 1.0 \times 10^{-1}$, $n_{\mathrm{amb}} = 1.0 \times 10^{-3}\ \mathrm{cm}^{-3}$ simulations, and its strength was only weakly anticorrelated with the initial distance between the clouds.

The clouds also change as they pass through the ambient medium. Ram pressure and shear constantly sweep back material from the front and sides of the clouds. Kelvin–Helmholtz instabilities pull lateral material into waves and curlicues. Low pressure develops inside the loops and behind dispersion components (e.g., Figure 15(b)), and a map of the number of components found along each line of sight (e.g., Figure 16). The component deconvolution reveals that some regions of high dispersion are due to multiple gas components along the line of sight. This can be seen by comparing the maps of the overall velocity dispersion (Figure 10), the velocity dispersion of the strongest component calculated using our version of GAUSSPY+ (Figure 14), and the number of velocity components (Figure 16). Furthermore, the region near $(x, y) = (-0.2,\ 0.0–0.2)$ in Figure 14 has been revealed to have a second or third component along the line of sight, distorting the overall dispersion seen in Figure 10. These second and third components are likely due to fragments that have been ablated from the clouds or accelerated from eddies generated around the low-pressure pocket behind the two clouds. In addition, the new figures more clearly reveal two lobes of lower-velocity material (approximately centered at $(x, y) = (0.1, 0.1)$ and $(x, y) = (-0.1, 0.1)$) in Figure 15(a) relative to Figure 15(b).

Lastly, while the multicomponent decomposition method was able to pick up accelerated fragments, it was unable to distinguish between the two clouds along the same line of sight. It treated both clouds as part of a single velocity component. In theory, this is due to the fact that both clouds should have roughly the same centroid velocity because they began with the same initial conditions.

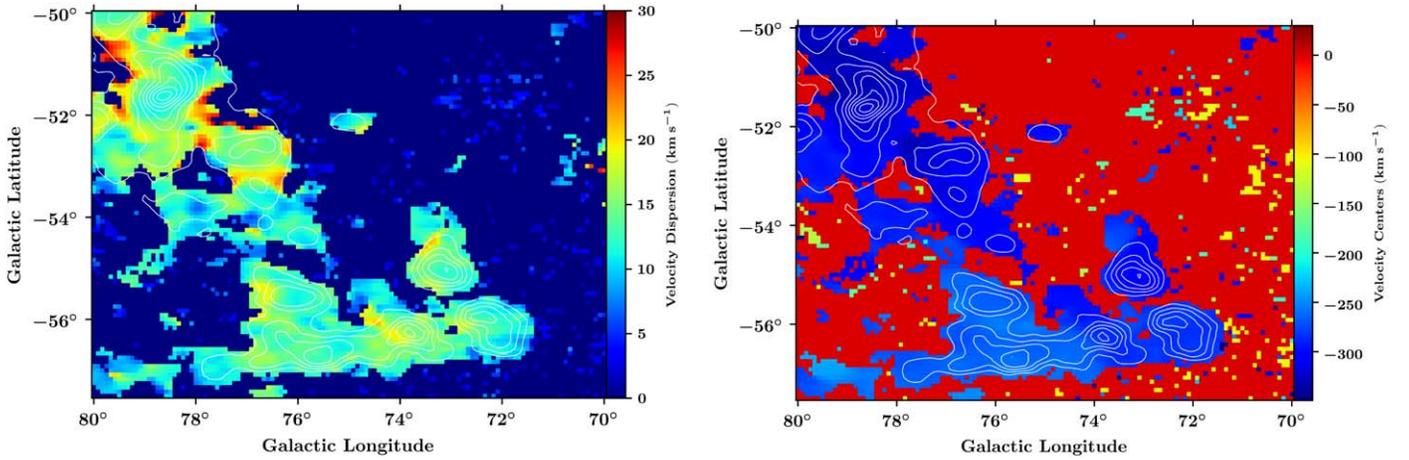

**Figure 13.** Map of the strongest component of velocity dispersion calculated using the multicomponent decomposition method (left) and map of the centroid velocity of the strongest component of velocity dispersion (right). The H I column density contours are overlaid in white on these data. This method was applied to the HI4PI data for the $l = (70.0, 80.0)$ and $b = (-50.0, -57.5)$ region of the sky.

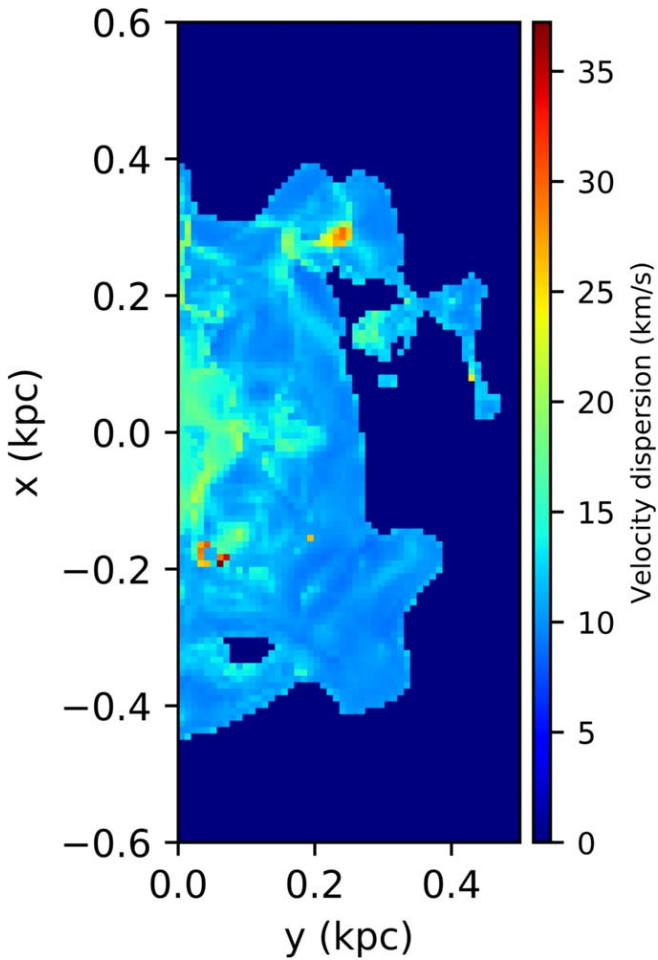

**Figure 14.** Velocity dispersion map of the strongest velocity component of the *dual_4r* simulation at 110 Myr viewed along the direction of motion of the clouds, calculated using the multicomponent decomposition method.

the clouds, driving more motion. Even in the interior, the material circulates. One result is that the clouds elongate to the extent that material from the leading cloud eventually overlaps with the trailing cloud. This occurs in all of our multicloud simulations, but the details depend on the density. In addition, the clouds break up easily in the lower-density simulations.

When they break up, fragments of cloud material are sent downwind and collide with the trailing clouds and cloud fragments.

The clouds exchange mass with the ambient medium. Ablation and evaporation of cloud gas competes with the entrainment and cooling of environmental gas in affecting the mass of the cloud. Condensation wins out in all of our simulations, but in some of the lower-density simulations, the competition is tilted in favor of evaporation for a period between ∼50 and ∼90 Myr before condensation ultimately wins (see Figure 4). During that era in our lower-density simulations, the more separated clouds suffer slightly more evaporation than the more closely clustered clouds. Therefore, in that era, the same simulations that retained more cloud mass were also the simulations with the greatest drafting. The most effective of our low-density simulations, the *dual_2rlow* simulation, condensed an additional three-sevenths of its mass before unrelated mass loss due to material flowing off of the domain began to exceed condensation at around 220 Myr. By comparison, the clouds in the higher-density simulations were more resistant to ablation and diffusion into the ambient gas. The ablation they did experience was mostly caused by Kelvin–Helmholtz instabilities that disrupted the tails of the clouds. In addition, the process of condensation was stronger in the higher-density set of simulations. In the most extreme case, the *dual_2r* simulation, the clouds were able to accrete close to three times their initial mass within 200–225 Myr, at which time cloud material began to flow out of the domain faster than ambient material accreted onto the cloud. Greater condensation in our higher-density simulations was expected, as the cloud masses were greater in the higher-density simulations than in the lower-density simulations, and previous hydrodynamical studies of condensation found that more massive clouds were more effective condensers of ambient material (Gritton et al. 2017).

We also examine the velocity dispersion of the material in the simulations and prepare several velocity dispersion maps. For each map, we select the simulated material whose temperature is no more than 10,000 K for better comparison with the neutral hydrogen seen by 21 cm observations. In some cases, higher dispersion is seen behind the trailing cloud, where streamlines of ambient material converge turbulently. There are generally higher median velocity dispersion values when the viewing perspective is along the direction of motion (as opposed to from the side). For example, when viewed along the direction of motion, as in Figure 10, the median dispersion is $\sigma = 11.6$ km s$^{-1}$. The material in Figure 5(b) has a median velocity dispersion of $\sigma = 9.9$ km s$^{-1}$. These values, from the *dual_4r* simulations at 110 Myr, are within $1\sigma$ of the median value of $\sigma = 10.6$ km s$^{-1}$ reported for individual clouds in the For et al. (2014) catalog.

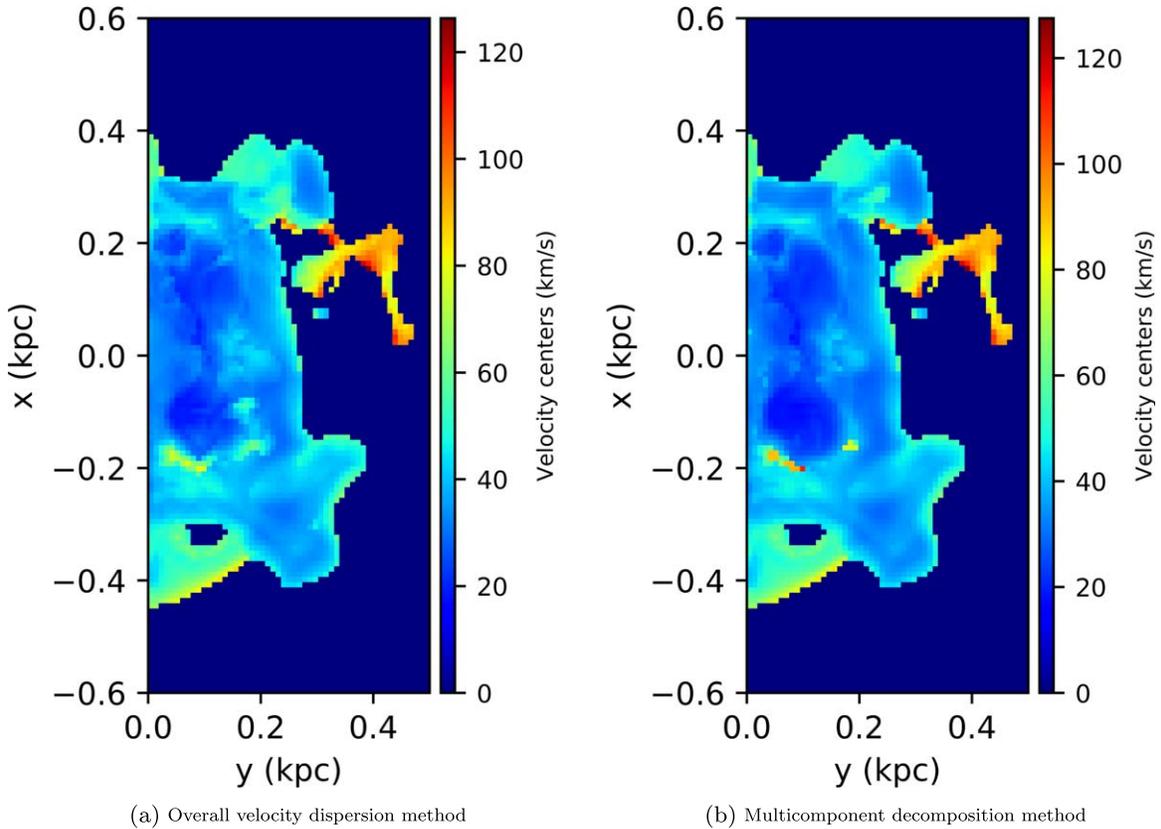

(a) Overall velocity dispersion method

(b) Multicomponent decomposition method

**Figure 15.** Velocity centroids map of the *dual_4r* simulation at 110 Myr from the perspective down the direction of flow of the clouds using both velocity dispersion methods. The velocity centroid values in panel (a) were calculated according to Equation (5). In panel (b), velocity centroid values are representative of the strongest component along each line of sight.

We also generated maps that show the velocity dispersion associated only with bulk motion, such as turbulence, velocity gradients, and overlapping material of different velocities. Practically speaking, these maps were generated by resetting the $T \leqslant 10^4$ K gas to a temperature of 0 K during the velocity dispersion calculation (see Figures 6, 17, and 18). For example, in the case of Figure 17 (which depicts the *dual_4r* simulation at 110 Myr), we found that most of the material has dispersion values due to motion of $\sigma \leqslant 3.0$ km s$^{-1}$, although there are some exceptional regions with much higher velocity dispersion due to motion. The same simulation viewed from the perspective along the direction of flow resulted in a median velocity dispersion value of $\sigma = 6.36$ km s$^{-1}$.

In the event that an astronomer would like a decomposition of the multiple velocity components to study, we have developed code that separates multiple overlapping components along the line of sight. We used this code to study the properties of the individual velocity components, create maps of the strongest velocity dispersion components, and determine the centroid velocities of the strongest components. We found that this tool enabled us to model more complex three-dimensional configurations of clouds, such as one cloud partially or entirely overlapping another cloud, and thus constrain other properties based on these simulations. For a cloud with fragments or another cloud along the same line of sight with it, this method more clearly resolves the velocity components of the core of the cloud, as seen in Figure 15, which can be verified by the map of the number of velocity components.

Cool temperatures have also been observed in MS clouds. A typical value is around 100 K (Matthews et al. 2009). When we use this value for the temperature of the clouds in our velocity dispersion maps, we find that the clouds appear as constant-value footprints with embedded core peaks. These footprints are similar to those visible in H I maps of the MS from the HI4PI survey data (HI4PI Collaboration et al. 2016) and could be used in the future to help constrain the core temperature of clouds of neutral hydrogen. The velocity dispersion in the footprint regions, near the extremities of the clouds, provides a better upper limit on the thermal broadening component and thus the temperature than the embedded higher-dispersion regions, as the latter could easily arise due to overlapping velocity components.

Our simulations indicate that drafting favors more rarefied clouds and backgrounds. Additionally, the clouds embedded in the lower of the two ambient densities experience different struggles with evaporation and condensation than their higher-density counterparts. The differences in the strengths of drafting and condensation between the lower-density simulations ($n_{cl} = 1.0 \times 10^{-2}$, $n_{amb} = 1.0 \times 10^{-4}$ cm$^{-3}$) and the higher-density simulations, which still model very rarefied regions ($n_{cl} = 1.0 \times 10^{-1}$, $n_{amb} = 1.0 \times 10^{-3}$ cm$^{-3}$), raise the question of which case is a better approximation of the MS environment. The lower cloud and environmental densities are more favored by the observations of Sembach et al. (2003), Grcevich & Putman (2009), Hsu et al. (2011), and Fox et al. (2014). These conditions are also favored by the models of MS IV (the fourth region of clouds along the MS identified by Mathewson et al. 1977), in that Murali (2000) found that the volume density of the extended halo and circumgalactic medium needs to be $n_{amb} \leqslant 10^{-5}$ cm$^{-3}$ in order for MS IV to survive on timescales of 500 Myr.

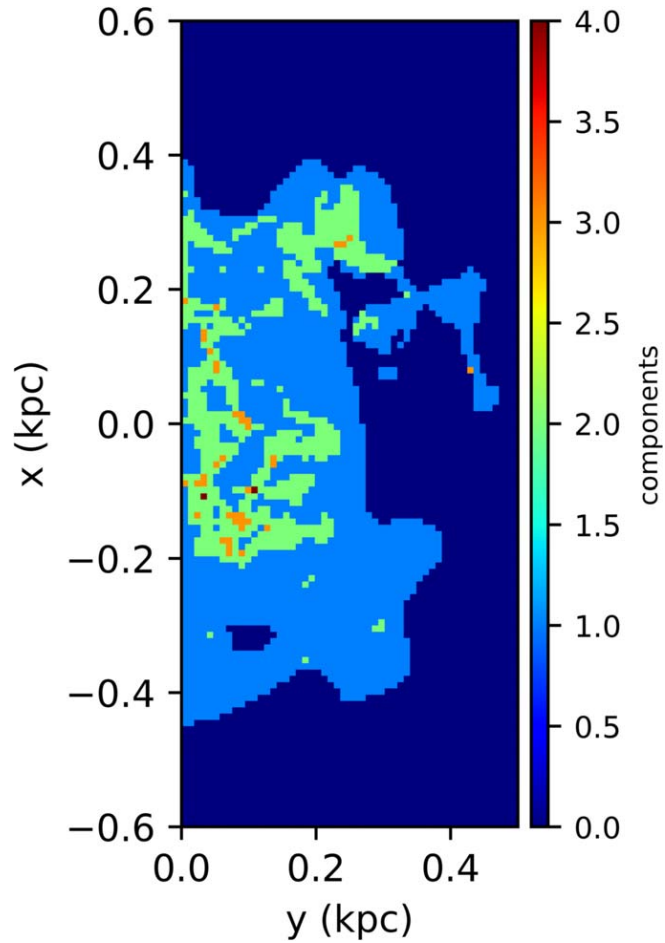

**Figure 16.** Map of the number of velocity components calculated by the multicomponent decomposition method for the *dual_4r* simulation at 110 Myr when viewed along the direction of the clouds' motion.

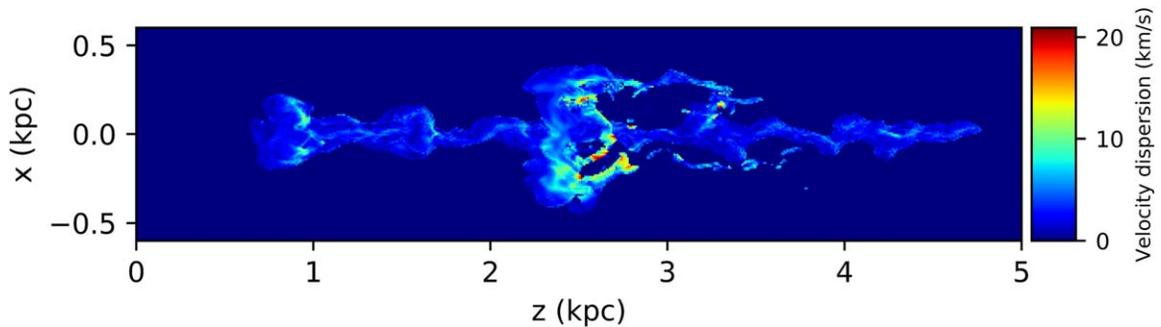

**Figure 17.** Velocity dispersion map of the *dual_4r* simulation at 110 Myr with the thermal broadening component removed.

The circumgalactic medium is not a welcoming environment for MS clouds. Estimates using Equation (2.4) from Bland-Hawthorn (2009) result in an upper limit on the distance that a typical cloud can travel before it is destroyed (velocity × lifetime) at 70 kpc for cloud–ambient density ratios up to 1000 and cloud diameters up to 2000 pc. Yet the MS stretches at least 200 kpc (D'Onghia & Fox 2016). How can clouds travel so far without diffusing into the ambient gas in the halo? Various papers have made efforts to show how clouds survive in the MS using mechanisms such as condensation (Vieser & Hensler 2007b, 2007a) and magnetic fields (McClure-Griffiths et al. 2010; Sander & Hensler 2019). We believe that adjacent cloud–cloud interactions are another contributing mechanism because multiple clouds assembled advantageously may provide drafting and mimic the dynamics of a larger cloud in offering protection for downstream clouds.

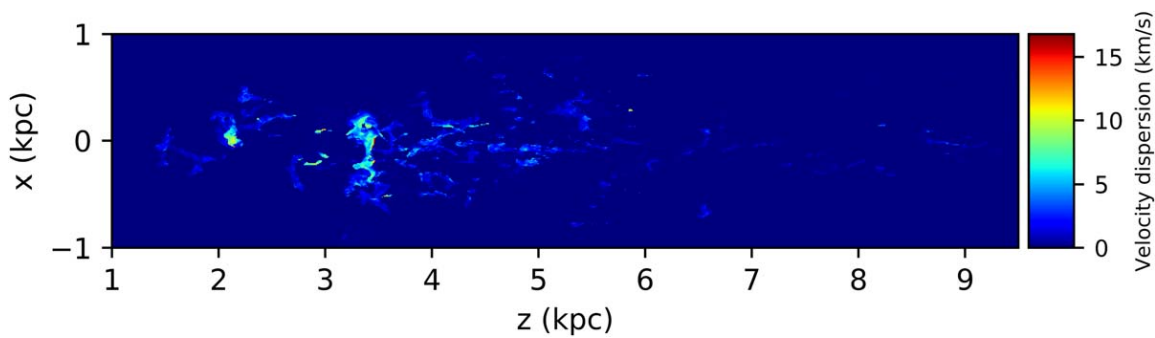

**Figure 18.** Velocity dispersion map of the *dual_4rlow* simulation at 200 Myr with the thermal broadening component removed.

We would like to thank Ashton (Rutkowski) Pond for her contributions to the method of tracking the clouds through the domain. The simulations were performed on the computer cluster at the Georgia Advanced Computing Resource Center (GACRC) at the University of Georgia. Shan-Ho Tsai at the GACRC provided invaluable help in troubleshooting hardware and software issues with the cluster. We would also like to thank the referee for their detailed and thoughtful comments to us during the review process. M.E.W. would like to thank the Center for Undergraduate Research (CURO) at The University of Georgia for funding during this project. R.L.S.'s contributions to this project were partially supported by grant No. NNX13AJ80G through the NASA ATP program. The Parkes Radio Telescope was used for the HI4PI survey. The Parkes Radio Telescope is part of the Australia Telescope National Facility which is funded by the Australian Government for operation as a National Facility managed by CSIRO. The EBHIS data are based on observations performed with the 100-m telescope of the MPIfR at Effelsberg. EBHIS was funded by the Deutsche Forschungsgemeinschaft (DFG) under the grants KE757/7-1 to 7-3.

**ORCID iDs**

M. Elliott Williams  https://orcid.org/0000-0001-8271-3394
Robin L. Shelton  https://orcid.org/0000-0001-5221-0315

**References**

Abuter, R., Amorim, A., Bauböck, M., et al. 2019, A&A, 625, L10

Bland-Hawthorn, J. 2009, in IAU Symp. 256, The Magellanic System: Stars, Gas, and Galaxies, ed. J. Andersen, J. Bland-Hawthorn, & B. Nordström (Cambridge: Cambridge Univ. Press), 489
Bland-Hawthorn, J., Sutherland, R., Agertz, O., & Moore, B. 2007, ApJ, 670, L109
Bregman, J. N., & Lloyd-Davies, E. J. 2007, ApJ, 669, 990
Brüns, C., Kerp, J., Staveley-Smith, L., et al. 2005, A&A, 432, 45
Chatard, J.-C., & Wilson, B. 2003, Med. Sci. Sports Exercise, 35, 1176
D'Onghia, E., & Fox, A. J. 2016, ARA&A, 54, 363
For, B. Q., Staveley-Smith, L., Matthews, D., & McClure-Griffiths, N. M. 2014, ApJ, 792, 43
Fox, A. J., Wakker, B. P., Barger, K. A., et al. 2014, ApJ, 787, 147
Fryxell, B., Olson, K., Ricker, P., et al. 2000, ApJS, 131, 273
Gatto, A., Fraternali, F., Read, J. I., et al. 2013, MNRAS, 433, 2749
Grcevich, J., & Putman, M. E. 2009, ApJ, 696, 385
Gritton, J. A., Shelton, R. L., & Galyardt, J. E. 2017, ApJ, 842, 102
Heitsch, F., & Putman, M. E. 2009, ApJ, 698, 1485
Henley, D. B., Shelton, R. L., & Kwak, K. 2014, ApJ, 791, 41
HI4PI Collaboration, Ben Bekhti, N., Flöer, L., et al. 2016, A&A, 594, A116
Hsu, W.-H., Putman, M. E., Heitsch, F., et al. 2011, AJ, 141, 57
Jin, S., & Lynden-Bell, D. 2008, MNRAS, 383, 1686
Kerp, J., Winkel, B., Ben Bekhti, N., Flöer, L., & Kalberla, P. M. W. 2011, AN, 332, 637
Kwak, K., Henley, D. B., & Shelton, R. L. 2011, ApJ, 739, 30
Mathewson, D., Schwarz, M., & Murray, J. 1977, PASA, 3, 133
Matthews, D., Staveley-Smith, L., Dyson, P., & Muller, E. 2009, ApJL, 691, L115
McClure-Griffiths, N. M., Madsen, G. J., Gaensler, B. M., McConnell, D., & Schnitzeler, D. H. F. M. 2010, ApJ, 725, 275
McClure-Griffiths, N. M., Pisano, D. J., Calabretta, M. R., et al. 2009, ApJS, 181, 398
Murali, C. 2000, ApJL, 529, L81
Putman, M. E., Peek, J. E. G., & Joung, M. R. 2012, ARA&A, 50, 491
Riener, M., Kainulainen, J., Henshaw, J. D., et al. 2019, A&A, 628, A78
Sander, B., & Hensler, G. 2019, MNRAS, 490, L52
Sembach, K. R., Wakker, B. P., Savage, B. D., et al. 2003, ApJS, 146, 165
Sutherland, R. S., & Dopita, M. A. 1993, ApJS, 88, 253
Vieser, W., & Hensler, G. 2007a, A&A, 472, 141
Vieser, W., & Hensler, G. 2007b, A&A, 475, 251
Westerweel, J., Aslan, K., Pennings, P., & Yilmaz, M. 2016, arXiv:1610.10082
Westmeier, T. 2017, MNRAS, 474, 289